\documentclass[10pt,a4paper,onecolumn]{IEEEtran}
\usepackage[lmargin=2cm, rmargin=2cm, tmargin=2cm, bmargin=2cm]{geometry}
\usepackage{ccaption}
\usepackage[T1]{fontenc}
\usepackage{graphicx}
\usepackage{amsfonts,amsmath,amsthm,bm,subfig,cite}

\DeclareMathOperator{\sat}{sat}

\title{Intelligent Sliding Mode Control of an Overhead Container Crane}
\author{Wallace Moreira Bessa, Svenja Otto, Edwin Kreuzer, Robert Seifried}
\date{}

\begin{document}

\maketitle

\abstract{

In this contribution, an intelligent controller is proposed for an underactuated overhead container crane subject to both parameter uncertainties and unmodeled dynamics. The adopted approach is based on the sliding mode method to confer robustness against modeling inaccuracies and external disturbances. Additionally, an adaptive fuzzy inference system is embedded within the control law to improve set-point regulation and trajectory tracking. In order to evaluate the performance of the proposed intelligent scheme, the control law was implemented and tested in a 1:6 scale experimental container crane, available at the Institute of Mechanics and Ocean Engineering at Hamburg University of Technology. The obtained experimental results demonstrate not only the feasibility of the proposed scheme, but also its improved efficacy for both stabilization and trajectory tracking problems.

}

\section*{Introduction}

Overhead container cranes are typical underactuated multibody systems since they have less independent control inputs than 
degrees of freedom. They play an essential role in cargo handling operations at ports and in the industrial sector. In order 
to ensure safe operating conditions, undesirable load swing should be avoided during the execution of a certain task.
In this case, a feedback control scheme could be adopted to automatically prevent oscillation of the load. However, the 
design of accurate controllers for this kind of problem can become very challenging inasmuch as the load swing cannot be 
controlled directly. Moreover, overhead crane dynamics are frequently uncertain, e.g.\ due to unknown friction forces, and 
are also subject to external disturbances such as wind loads.

Intelligent control has proven to be a very attractive approach to cope with uncertain nonlinear systems 
\cite{tese,cobem2005,Bessa2017,Bessa2018,Bessa2019,Deodato2019,Lima2018,Lima2020,Lima2021,Tanaka2013}. 
By combining nonlinear control techniques, such as feedback linearization or sliding modes, with 
adaptive intelligent algorithms, for example fuzzy logic or artificial neural networks, the resulting 
intelligent control strategies can deal with the nonlinear characteristics as well as with modeling 
imprecisions and external disturbances that can arise.

In this contribution, an intelligent controller is proposed for an underactuated overhead container crane subject to both parameter 
uncertainties and unmodeled dynamics. The adopted approach is based on the sliding mode method to confer robustness against modeling 
inaccuracies and external disturbances. Additionally, an adaptive fuzzy inference system is embedded within the control law to improve 
set-point regulation and trajectory tracking. In order to evaluate the performance of the proposed intelligent scheme, the control law 
was implemented and tested in a 1:6 scale experimental container crane, available at the Institute of Mechanics and Ocean Engineering 
at Hamburg University of Technology. The experimental setup, as shown in Figure~\ref{fig:setup}, consists of a trolley and a container 
with dimensions $0.35\,\mathrm{m}\times0.37\,\mathrm{m}\times0.86\,\mathrm{m}$ that is attached to the trolley by four cables.

\section*{Controlling the container crane}

Since the trolley moves along only a linear axis, planar motion is assumed, yielding the equation of motion:

\begin{equation}
\left[ \begin{array}{ccc}
M+m		&m\sin\theta	&ml\cos\theta\\
m\sin\theta	&m		&0\\
ml\cos\theta	&0	&ml^2 \end{array} \right]
\left[ \begin{array}{c} \ddot{x}\\ \ddot{l}\\ \ddot{\theta} \end{array} \right]=
\left[ \begin{array}{c} m\dot{\theta}(\dot{\theta}l\sin\theta-2\dot{l}\cos\theta)\\ 
			m(l\dot{\theta}^2+g\cos\theta)\\
			-ml(2\dot{l}\dot{\theta}+g\sin\theta) \end{array} \right]+
\left[ \begin{array}{c} u_x\\ u_l\\ 0 \end{array} \right].
\label{eq:crane}
\end{equation}

Here, $u_x$ and $u_l$ are, respectively, the control forces acting on the trolley and the cables, $x$ is the trolley 
position, $l$ stands for the length of the cables, $\theta$ represents the swing angle, $M$ is the mass of the trolley, 
and $m$ is the container mass. It should be emphasized that only the variables $x$ and $l$ can be directly actuated. 
The swing angle $\theta$, on the other hand, is an unactuated variable.

Now, following the sliding mode method, a stable manifold should be established in the state space. Here, the switching 
variables are defined according to the general approach proposed by Ashrafiuon and Erwin \cite{Ashrafiuon2008}:

\begin{equation}
\mathbf{s}=\left[ \begin{array}{cc}
\alpha_x	&0\\
0		&\alpha_l \end{array} \right]
\left[ \begin{array}{c} \dot{\tilde{x}}\\ \dot{\tilde{l}}  \end{array} \right]+
\left[ \begin{array}{cc}
\lambda_x	&0\\
0		&\lambda_l \end{array} \right]
\left[ \begin{array}{c} \tilde{x}\\ \tilde{l}  \end{array} \right]+
\left[ \begin{array}{c} \alpha_\theta\\ 0  \end{array} \right]\dot{\tilde{\theta}}+
\left[ \begin{array}{c} \lambda_\theta\\ 0  \end{array} \right]\tilde{\theta}.
\label{eq:surf}
\end{equation}

Here, $\mathbf{s}=[s_x\ s_l]^\top$ is the switching vector, $\alpha_n$ and $\lambda_n$ (with $n=x,l,\theta$) are parameters 
that must be properly chosen in order to ensure the stability of the sliding surface \cite{Ashrafiuon2008}, and $\tilde{x}
=x-x_d$, $\tilde{l}=l-l_d$, and $\tilde{\theta}=\theta-\theta_d$ are the tracking errors, with subscript $d$ representing 
the desired values for each variable.

Conventionally, sliding mode based schemes are implemented with a discontinuous term in the control law, but a known drawback 
of this approach is the resulting chattering phenomenon. In order to avoid the undesired effects of the control chattering, 
a thin boundary layer neighboring the switching surface could be adopted, by replacing the discontinuity with a continuous 
interpolation inside the boundary layer. This substitution can minimize or, when desired, even completely eliminate chattering. 
However, it turns a perfect tracking into a tracking with guaranteed precision problem, which actually means that a steady-state 
error will always remain.

Thus, in order to enhance the control performance, an adaptive fuzzy compensator $\mathbf{\hat{d}}=[\hat{d}_x\ \hat{d}_l]^\top$ 
is combined with a smooth sliding mode controller:

\begin{multline}
\mathbf{u}=-\left[ \begin{array}{cc}
-(\alpha_\theta\cos\theta-\alpha_xl)/Ml	&(\alpha_\theta\cos\theta-\alpha_xl)\sin\theta/Ml\\
-\alpha_l\sin\theta/M		&\alpha_l(m\sin^2\theta+M)/Mm \end{array} \right]^{-1}
\bigg[ \begin{array}{l} 
\alpha_\theta(2\dot{l}\dot{\theta}+g\sin\theta)/l+\\ 
\alpha_l(l\dot{\theta}^2+g\cos\theta)+ 
\end{array}\\
\begin{array}{r} 
+\hat{d}_x-\alpha_x\ddot{x}_d-\alpha_\theta\ddot{\theta}_d+\lambda_x\dot{\tilde{x}}+\lambda_\theta\dot{\tilde{\theta}}
+K_x\sat(s_x/\phi_x)\\ 
+\hat{d}_l-\alpha_l\ddot{l}_d-\lambda_l\dot{\tilde{l}}+K_l\sat(s_l/\phi_l)
\end{array} \bigg],
\label{eq:ismc}
\end{multline}

\noindent
where $\mathbf{u}=[u_x\ u_l]^\top$ is the proposed control signal, $K_x$ and $K_l$ are the related control gains, $\sat(\cdot)$ is
the standard saturation function, and $\phi_x$ and $\phi_l$ define the width of the associated boundary layers.

Each compensation term is computed using the Takagi-Sugeno-Kang inference method \cite{jang1}: $\hat{d}_m=\mathbf{\hat{D}}^\top_m
\bm{\Psi}_m$ (with $m=x,l$), where $\mathbf{\hat{D}}_m=[\hat{D}_{m1},\hat{D}_{m2},\dots,\hat{D}_{mR}]^\top$ are vectors containing 
the attributed values $\hat{D}_{mr}$ to each rule $r$, $\bm{\Psi}_m=[\psi_{m1},\psi_{m2},\dots,\psi_{mR}]^\top$ are vectors with 
components $\psi_{mr}=w_{mr}/\sum_{r=1}^{R}w_{mr}$ and $w_{mr}$ is the firing strength of each rule. With the view to reduce the 
number of fuzzy sets and rules, the switching variables $s_m$, instead of all state variables, are considered in the premise of the 
related fuzzy rules. The vectors of adjustable parameters are automatically updated by the adaptation laws $\mathbf{\dot{\hat{D}}}_m
=\varphi_m s_m\bm{\Psi}_m$, where $\varphi_m$ are strictly positive constants related to the adaptation rate \cite{rsba2010}. 

\begin{figure}[htb]
\centering
\mbox{
\subfloat[Experimental setup: trolley, cables and container.]{\label{fig:setup}
\includegraphics[width=0.4\textwidth]{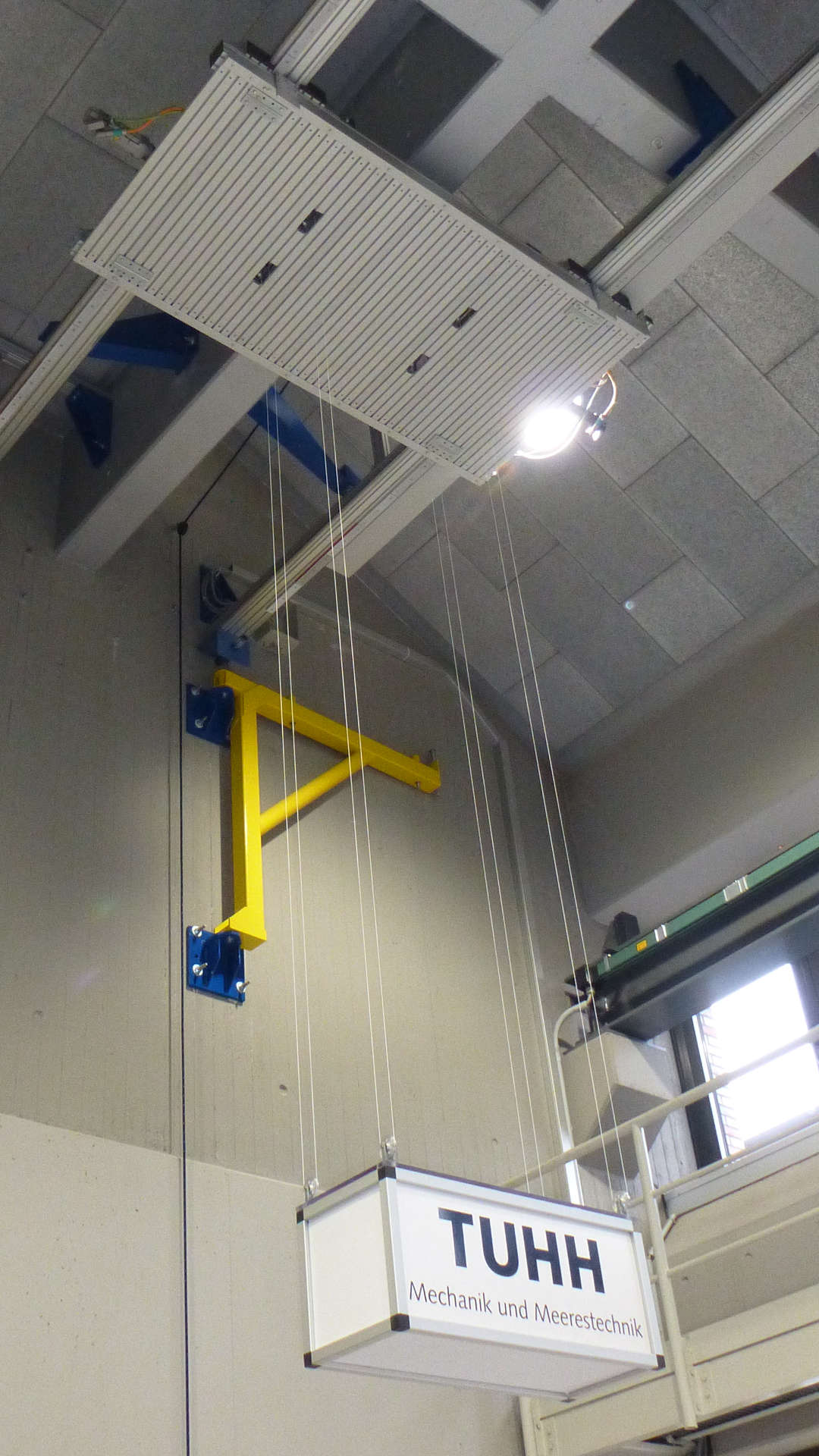}}
\quad\quad
\subfloat[Tracking of a semicircle around an obstacle.]{\label{fig:semicircle}
\includegraphics[width=0.4\textwidth]{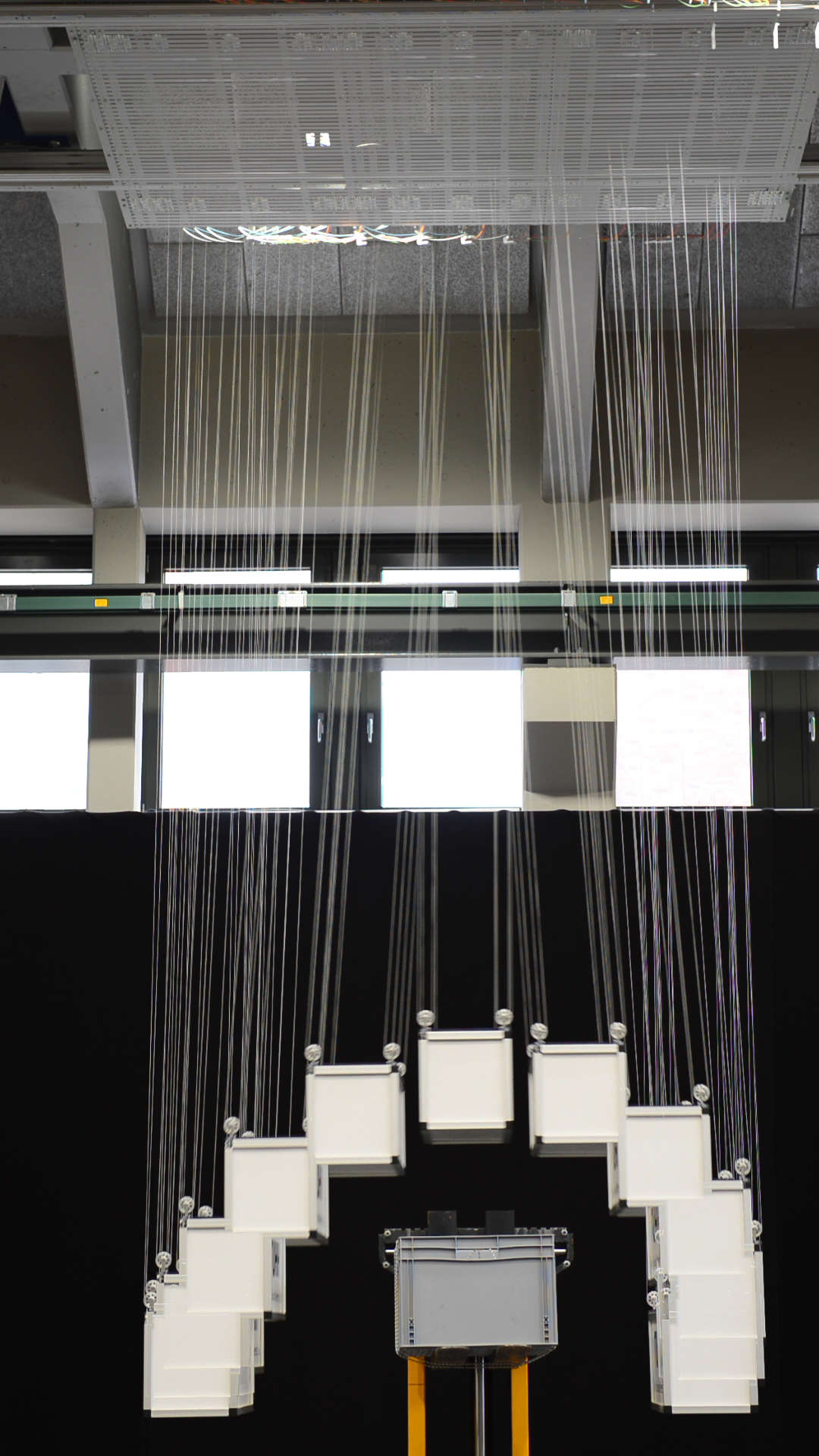}}
}
\caption{Overhead container crane at the Institute of Mechanics and Ocean Engineering.}
\label{fig:crane}
\end{figure}

Finally, the proposed intelligent controller is evaluated in the experimental overhead container crane for both stabilization and 
trajectory tracking problems. Figure~\ref{fig:semicircle} shows, for example, the overlayed video frames related to the tracking of 
a semicircle trajectory around an obstacle.

Due to the ability of the adopted adaptive fuzzy algorithm to recognize and compensate for unmodeled dynamics, e.g.\ friction 
or damping forces, the improved performance of the proposed intelligent sliding mode controller over the conventional counterpart 
could be clearly verified in all conducted experiments.

\end{document}